\documentclass[referee,sn-basic]{sn-jnl}

\usepackage{graphicx}%
\usepackage{multirow}%
\usepackage{amsmath,amssymb,amsfonts}%
\usepackage{amsthm}%
\usepackage{mathrsfs}%
\usepackage[title]{appendix}%
\usepackage{xcolor}%
\usepackage{textcomp}%
\usepackage{manyfoot}%
\usepackage{booktabs}%
\usepackage{algorithm}%
\usepackage{algorithmicx}%
\usepackage{algpseudocode}%
\usepackage{listings}%

\theoremstyle{thmstyleone}%
\theoremstyle{thmstyletwo}%

\theoremstyle{thmstylethree}%

\begin{document}

\title[Article Title]{CorStitch: Accessible Video Stitching for Coral Monitoring}


\author*[1]{\fnm{Julian Christopher} \sur{Maypa}}\email{jmaypa@nip.upd.edu.ph}

\author[1]{\fnm{Johnenn} \sur{Manalang}}\email{jrmanalang1@up.edu.ph}
\equalcont{These authors contributed equally to this work.}

\author[2]{\fnm{Wilfredo} \sur{Licuanan}}\email{wilfredo.licuanan@dlsu.edu.ph}
\equalcont{These authors contributed equally to this work.}

\author[1]{\fnm{Maricor} \sur{Soriano}}\email{jing@nip.upd.edu.ph}
\equalcont{These authors contributed equally to this work.}

\affil*[1]{\orgdiv{National Institute of Physics}, \orgname{University of the Philippines Diliman}, \orgaddress{\city{Quezon City}, \country{Philippines}}}

\affil[2]{\orgdiv{Department of Biology}, \orgname{De La Salle University}, \orgaddress{\street{Taft Avenue}, \city{Manila}, \country{Philippines}}}


\abstract{We develop CorStitch, a video-stitching software that automates the process of stitching underwater videos for coral reef monitoring and assessment. It offers a free, flexible, and user-friendly alternative to existing stitching software that may be difficult to access. CorStitch utilizes a Fourier-based image registration algorithm to stitch the central horizontal strips of successive frames of down-looking belt and dive transect videos, generating georeferenced and marked mosaics. Tests show that CorStitch can produce high-quality mosaics that are comparable to those generated by existing stitching software, with the added advantage of being open-source and accessible to users with varying levels of technical expertise. CorStitch has the potential to supplement the coral reef monitoring efforts of local communities, government agencies, and non-governmental organizations.
}




\maketitle
\section{Introduction}\label{sec:Intro}
Affordable underwater cameras have made down-looking underwater video transects a popular method for monitoring and assessing coral reefs. Underwater videos provide permanent visual records of the condition of coral reefs, which can be further processed to extract benthic distribution \citep{hadi}, monitor changes in coral reefs \citep{monitoring}, create 3D imagery using photogrammetry \citep{photogrammetry, photogrammetry1}, and validate habitat maps generated by remote sensing imagery and classification \citep{remote, remote1}. With the increasing demand for coral reef assessments from local governments and non-governmental organizations \citep{edmunds}, developing a fast, efficient, and accessible workflow in processing underwater transect videos across varying spatial scales would significantly help reef monitoring efforts.

In the Philippines, there are two video-based methods that are used in reef monitoring. The Automated Rapid Reef Assessment System (ARRAS) was a program funded by the Department of Science and Technology (DOST) from 2010 to 2012, which aimed to develop technology to hasten the capture and processing of kilometer-long underwater belt transects \citep{shoestring}. ARRAS features a boat-towable platform where a waterproof camera \citep{teardrop} and a GPS echosounder can be installed. The data processing side of ARRAS involves a stitching software to convert videos into georeferenced mosaics \citep{corpuz}. In 2019, the Biodiversity Management Bureau of the Department of Natural Resources (DENR-BMB) in its Technical Bulletin No. 2019-04 \citep{bmbbulletin} recommended the use of ARRAS for Coral Reef Habitat and Associated Reef Fishes Assessment, and for reconnaissance and mapping purposes. The latest stitching and georeferencing software of ARRAS, known as KikoMaan, was licensed by UP Diliman in 2021 to Antipara Exploration, a company that provides underwater mapping and analytics services.

While ARRAS has facilitated large-scale reef mapping, diver-operated video transects remain the primary survey method used by many local government units, non-governmental organizations, and academic monitoring programs \citep{C30Licuanan, C5_Licuanan}. Consequently, image processing tools intended to support community-based reef monitoring should be compatible with both towed belt transects and diver-operated transects. Effective community-based reef monitoring also requires tools that are compatible with the survey methods already employed by local government units and non-governmental organizations \citep{CS_Alcala_Russ,gouraguine}. Since both ARRAS belt transects and C5 dive transects are used in the Philippines, image processing software should ideally support both acquisition methods.

Commercially available stitching software such as Agisoft Metashape \citep{agisoftwebsite} has been a popular choice for monitoring coral reefs \citep{noaa2019, burns2015, casoli2021} due to its ability to generate high-quality photomosaics. There are also other stitching programs that are freely available to the public, such as Autostitch \citep{autostitch} and mosaic-library \citep{mosaic_lib}. Despite the convenience that these programs provide, there may be issues when deploying them to a large number of independent users. Small fishing communities and municipalities may not have the budget to purchase software that requires a paid license. Additionally, stitching software that is not specifically designed to process video transect data may incorrectly stitch the frames together, as demonstrated by Aguinaldo and Soriano \citep{paaronama}. Thus, there is value in a supplementary open-source stitching software specifically designed for underwater transect videos that can complement existing commercial and open-source tools.

To supplement existing software and provide an option for users on a limited budget, we developed CorStitch, an open-source stitching software designed to generate mosaics from down-looking underwater video transects acquired using either towed camera systems or diver-operated surveys. We aim to provide a free, flexible, and user-friendly software that can supplement the coral reef monitoring efforts of local communities, government agencies, and non-governmental organizations. CorStitch is designed to be accessible to users with varying levels of technical expertise, allowing the end users to create mosaics with minimal training and equipment. In line with this, we demonstrate that CorStitch can produce high-quality mosaics that can:

\begin{itemize}

    \item aid in the monitoring and assessment of coral reefs by providing a visual record of the seafloor that can be used to identify benthos along a belt or dive transect,

    \item serve as a georeferenced record of the seafloor that can be used to track changes in the distribution of benthic biota over time, and

    \item facilitate expert review and verification by providing a permanent visual record that can be revisited for quality control, annotation, and independent assessment.

\end{itemize}

Succeeding sections are organized as follows: Section 2 discusses the related work that served as the basis for the workflow of CorStitch. Section 3 details the workflow and stitching process. Section 4 highlights CorStitch's video mosaicking capabilities, demonstrating effectiveness for both dive and belt transect videos. Section 5 discusses the features of CorStitch and their relevance to creating mosaics. Lastly, Section 6 summarizes our findings and recommendations.

\section{Related Work}



Underwater image mosaicking has become an important tool for processing diver-acquired photographs and video transects because it converts large collections of overlapping images into continuous representations of the seafloor. One of the commonly used diver-based methods of monitoring coral reefs is RGB imaging. In RGB imaging, divers will take photos or videos of the seafloor along a survey area, such as a transect or photo quadrat. The main advantage of RGB imaging is that it reduces the diver's ``bottom time'' since the data can be processed in the laboratory \citep{Current_methods}. The images and videos can be processed into mosaics by using feature-based methods. Several feature-based approaches have been proposed for underwater mosaicking. These include Harris corner-based registration \citep{stitch1}, improved SIFT methods for handling blurry underwater imagery \citep{stitch2}, and trajectory-aware mosaicking algorithms that discard poorly aligned frames prior to stitching \citep{stitch3}. The mosaics can then be further processed by machine learning models such as CoralNet \citep{CoralNet}, ReefCloud \citep{ReefCloud}, and SCoralDet \citep{SCoralDet} for automatic coral segmentation and analysis.

Feature-based image registration has been widely adopted for underwater image mosaicking because of its robustness in matching overlapping images. An alternative approach is featureless image registration, which estimates image alignment directly from image intensities rather than from extracted keypoints. Among these methods, Fourier-based image registration has been successfully applied to underwater video mosaicking and forms the basis of the approaches developed by Corpuz et al. \citep{corpuz} and Aguinaldo and Soriano \citep{paaronama}.

The work of Corpuz et al. \citep{corpuz} used a featureless warping method in creating video mosaics. By calculating the correlation peak between two frames, they were able to create 6-second mosaics using 180 video frames. Because the mosaics were generated from non-adjacent frames, parallax-induced misalignment became more pronounced. The work of Aguinaldo and Soriano \citep{paaronama} created the software named Paaronama. Their software is a video mosaicking algorithm that reduces parallax-induced errors by minimizing the angular field of view. By using image registration similar to \citep{corpuz}, they were able to create video mosaics that outperformed commercially available mosaicking software such as AutoStitch \citep{autostitch} and Microsoft Image Composite Editor. 

Despite advances in underwater image mosaicking, there remains value in an open-source software package that combines efficient featureless registration with an accessible workflow for underwater video transects.  Unlike previous implementations that focused primarily on the underlying mosaicking algorithm, CorStitch integrates mosaicking, georeferencing, and random point overlay into a single open-source software package that supports both towed and diver-operated underwater video transects.

\section{Methodology}\label{sec:Methods}

\subsection{CorStitch Processing Workflow}
The workflow of CorStitch is inspired by the methods used by Paaronama \citep{paaronama}. While Paaronama focused primarily on video mosaicking, CorStitch extends the workflow by integrating georeferencing, random point overlay, and an accessible graphical interface into a unified software package. The primary output of CorStitch is sets of video mosaics, which are generated by stitching information from successive frames together using a Fourier-based image registration algorithm. This section describes the CorStitch framework and the process of generating mosaics from transect videos.

\begin{figure}[h]
\centering
\includegraphics[width=0.99\textwidth]{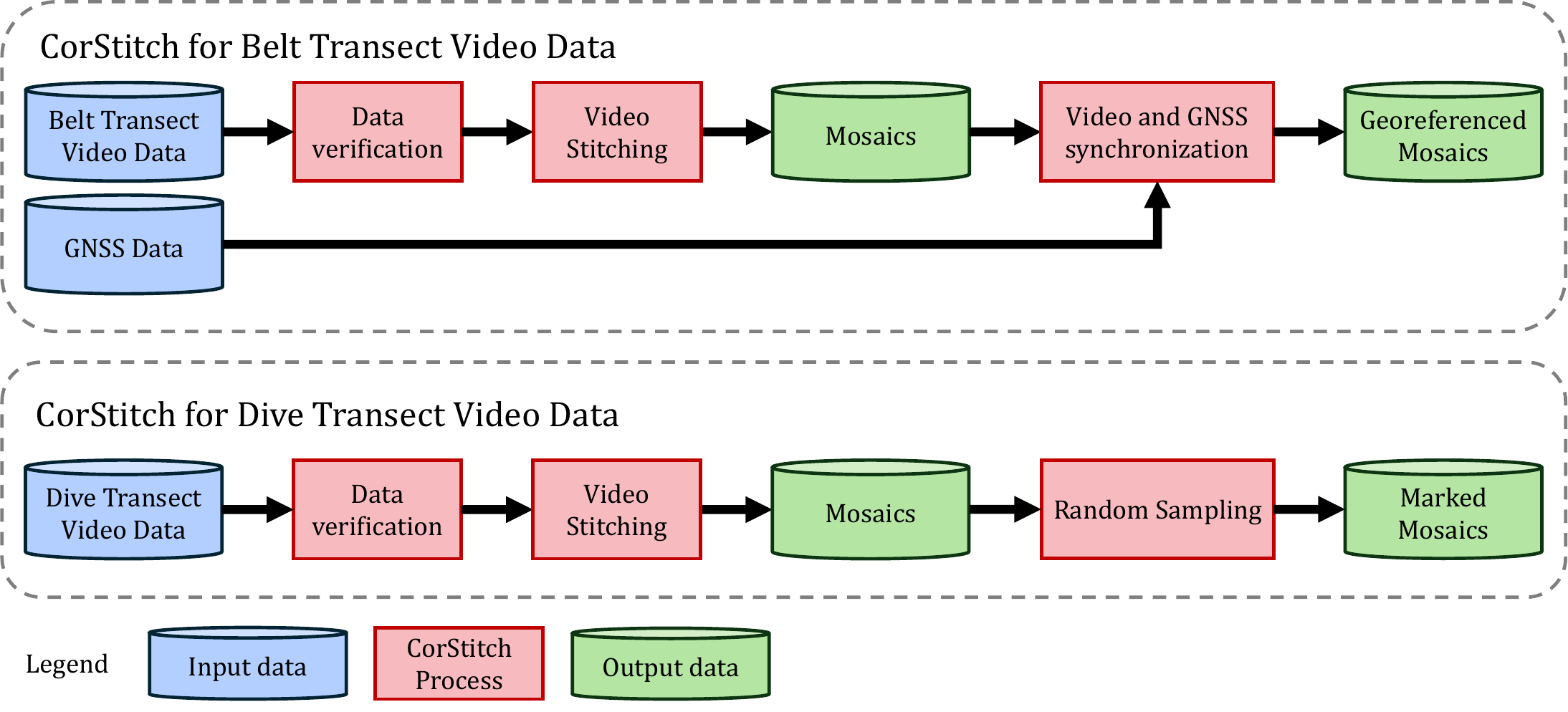}
\caption{CorStitch workflow showing the processing pipeline for belt transect and dive transect video data.}\label{fig:workflow}
\end{figure}

Figure \ref{fig:workflow} shows the overall processing workflow of CorStitch for both belt transect and dive transect video data. The first main process is video stitching, where information from each frame is extracted and stitched together to create a mosaic. These mosaics can be further processed into georeferenced and marked mosaics for belt transect and dive transect video data, respectively. Georeferenced mosaics are made by assigning Global Navigation Satellite System (GNSS) coordinates to each mosaic, which can be viewed in geospatial visualization tools such as Google Earth. As for the marked mosaics, they are made by overlaying randomly spaced markers in the mosaic to simulate the sampling process of CPCe \citep{cpce}.

\subsection{Underwater video mosaicking}
CorStitch is designed to process underwater videos, thus, spatial distortions and color imbalances are expected. Although the current version of CorStitch will not apply any corrections to stay true to our lightweight design, we made approximations and simplifications to take advantage of these underwater phenomena.

The color imbalance in underwater videos is due to the attenuation of light in water. The water column scatters red light the most, then green, and then blue. This gives the video a predominantly blue or green tint when there is no white balancing. The tint indicates that most of the information is stored in the blue or green channel. Thus, we only used the green channel to find the correlation peak. 

Spatial distortion can heavily impact frame correlation. Due to the planar shift of the camera, an object in $f$ will have a different magnification in $g$ due to the changing object-to-camera distance. This difference in magnification, which is a parallax error, can affect the location of the correlation peak if the camera is towed at high speeds. To combat this effect, we minimize the angular field of view (FOV) by only using the horizontal central strip of each frame. Minimizing the angular FOV approximates the image to a telecentric image, thus reducing the parallax errors.

During our experiments, we noticed that CC fails to correlate images with large patches of sand. This is because sand patches have high brightness, thus giving inaccurately high correlation values \citep{cross_correlation}. Also, PC fails when the features are distorted (e.g., motion blur), while CC does not. This could be due to the varying brightness of the features, which aids in the correlation of CC. With this, CorStitch uses both CC and PC to correlate $f$ and $g$. We set PC to be the accepted value, but when the displacement of the correlation peak from the center is more than $20\%$ of the strip's side length, the value obtained from CC is used instead. 

\subsection{Mosaic Creation Workflow}

\begin{figure}[h]
\centering
\includegraphics[width=0.99\textwidth]{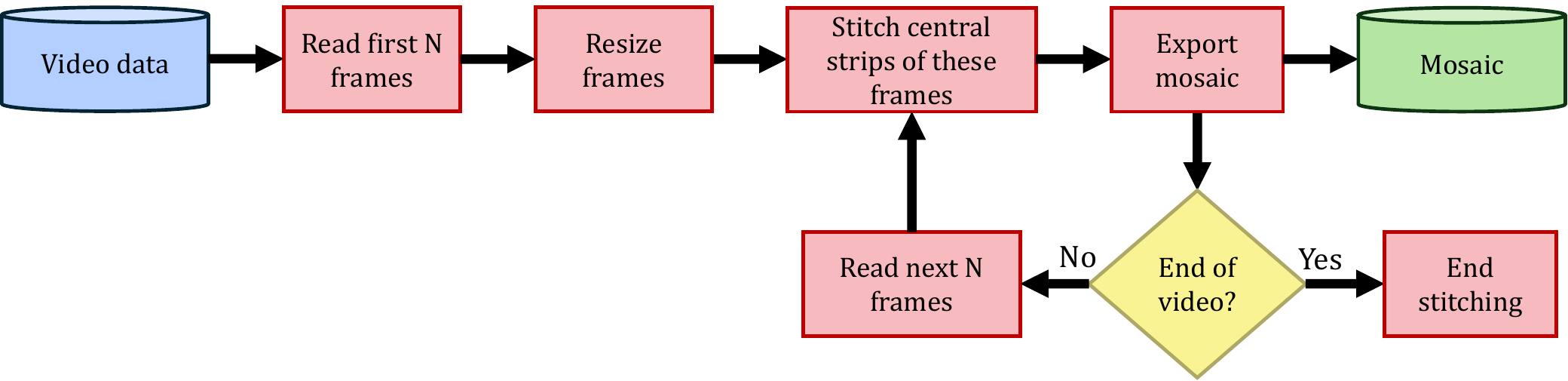}
\caption{The mosaic creation workflow of CorStitch.}\label{fig:stitch_flow}
\end{figure}

With the adjustments made to process the underwater videos, the mosaics are made by following the stitching workflow illustrated in Figure \ref{fig:stitch_flow}.

From the raw video data, the first $N$ frames are read. The value $N$ is obtained by multiplying the video frame rate by a user-specified mosaic time $t$. This means that mosaics will be made up of $t$-seconds worth of frames, and the total number of mosaics can be determined by dividing the length of the raw video by the mosaic time. Specifying the mosaic time is only available for processing belt transect videos. For the dive transect version, the whole video is stitched together to form one mosaic.

After getting the subset of frames, they are then resized to $480\,\mathrm{px} \times 360\,\mathrm{px}$, $854\,\mathrm{px} \times 480\,\mathrm{px}$, $1280\,\mathrm{px} \times 720\,\mathrm{px}$, or $1920\,\mathrm{px}\times 1080\,\mathrm{px}$, which are also known as 320p, 480p, 720p, and 1080p, respectively. This option allows the user to downscale their video frames before creating the mosaics. This can be beneficial to users who have hardware limitations. The needed memory and computing power to create these mosaics scale with the mosaic resolution, thus, lowering the resolution allows low-end computers to create mosaics.

\begin{figure}[h]
\centering
\includegraphics[width=0.99\textwidth]{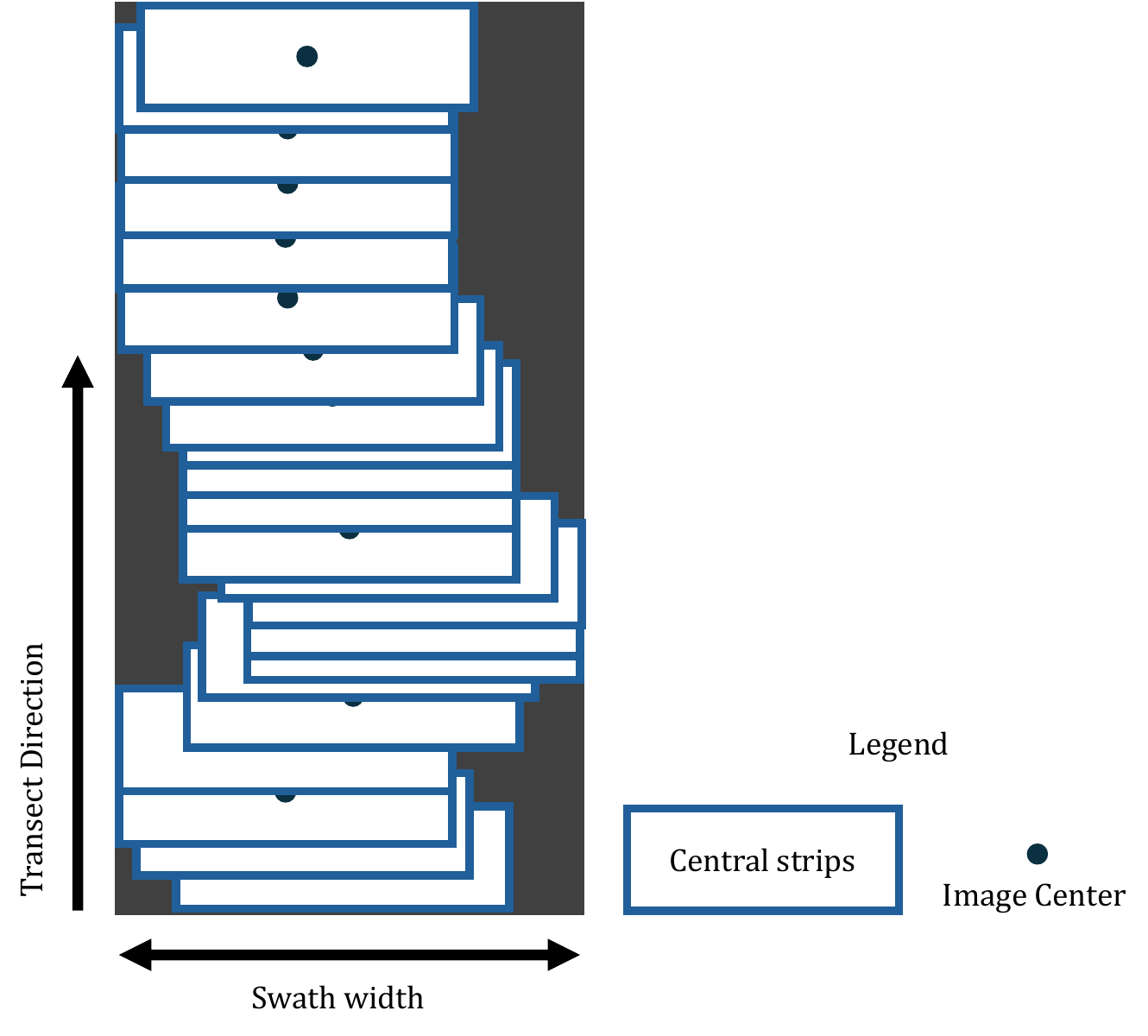}
\caption{The general composition of each mosaic.}\label{fig:mosaic_comp}
\end{figure}

Once the frames have been resized, the central strips of each frame are stitched together and exported into mosaics. This process repeats until all of the frames of the video are used. Using this workflow, each mosaic will generally have a composition shown in Figure \ref{fig:mosaic_comp}. The mosaic extends along the \textit{transect direction}, or the along-track direction. Meanwhile, the width of the mosaic is defined to be the \textit{swath width}, which is also known as the across-track direction.

\subsection{Stitching Algorithm}
It is assumed that the underwater transect videos are acquired with the camera maintained in a downward-facing orientation while moving along the transect. As a result, consecutive frames generally exhibit high overlap, and the relative motion between frames is expected to be dominated by translation, with only small rotations and scale changes. CorStitch employs Fourier-based image registration because these acquisition conditions are well suited to phase correlation, which provides a direct estimate of image translation without requiring feature detection and descriptor matching.

In the process of converting the video data to mosaics, we employ a stitching algorithm that correlates the two central strips of successive frames. If two image functions $f$ and $g$ correspond to two adjacent strips that are related by a planar shift, then we write $g$ as 
\begin{equation}\label{eq:relfg}
    g(x,y) = f(x + \Delta x,\, y + \Delta y)
\end{equation}
where $\Delta x$ and $\Delta y$ correspond to the horizontal and vertical shifts, respectively. The overlap between $f$ and $g$ is then obtained by getting the cross-correlation
\begin{equation}\label{eq:CC}
    \mathrm{CC}(x,y) = \mathcal{F}^{-1}\left\{F(u,v)^\ast \circ G(u,v)\right\},
\end{equation}
and phase-correlation
\begin{equation}\label{eq:PC}
    \mathrm{PC}(x,y) = \mathcal{F}^{-1}\left\{ \frac{F(u,v)^\ast \circ G(u,v)}{\left|F(u,v)^\ast \circ G(u,v)\right|}\right\} =\mathcal{F}^{-1}\left\{ \exp(-i(u\Delta x + v\Delta y))\right\},
\end{equation}
between each image. The superscript $^\ast$ denotes the complex conjugate of the function, $\circ$ is the Hadamard product, $\mathcal{F^{-1}}$ is the 2D inverse Fourier Transform, and $F$ and $G$ are the Fourier transforms of $f$ and $g$, respectively \citep{cross_correlation}. The maximum value, or the peak, of CC and PC is the point at which the two images overlap the most with respect to the center of the $f$. Given the relation in \ref{eq:relfg}, the correlation peak of $f$ and $g$ is displaced from the center by $\Delta x$ horizontally and $\Delta y$ vertically.  By overlapping $f$ and $g$ based on the correlation peak, the two images can be stitched together to create a visually continuous mosaic.

Because CorStitch estimates only translational displacement between successive frames, the registration procedure remains computationally lightweight. However, non-planar translations such as elevation changes, rotations, and perspective changes cannot be calculated. Thus, excessive non-planar movements will result in stitching errors that will lower the quality of the underwater video mosaic.

\subsection{Georeferenced Mosaics}
When GNSS data is available, CorStitch can georeference the mosaics by synchronizing the local time with the video time. The synchronization is done by the \textit{call out} method. To perform a call out, we verbally state the date, time, and location, and show the GNSS clock to the camera before submerging the camera underwater. By showing the GNSS clock to the camera, we can synchronize the local time with the video time, which in turn allows us to synchronize the longitudes and latitudes with the video time.

Using the longitude, latitude, and bearing data, we use Python's \texttt{SimpleKML} to assign GNSS coordinates to the corners of each mosaic. If bearing data is unavailable, CorStitch uses the Haversine formula to calculate the bearing from the given longitude and latitude data. The georeferenced mosaics are compiled and exported into zipped Keyhole Markup Language (KMZ) files, which can be viewed in geospatial visualization tools.

\subsection{Marked Mosaics}
The process of marking the mosaics is similar to the process of marking in CPCe. Pixels in the mosaic are chosen using uniform random sampling. A cross mark and a number are overlaid on each chosen pixel. The user can then proceed to mark the mosaics by identifying the benthos under each cross mark.

Although CorStitch uses uniform random sampling to produce marked mosaics, there is a probability that the markings will form small clusters. These small clusters will result in a biased sampling of the benthic distribution. To avoid this, we prevent points from being sampled if they are within a certain radius of a previously sampled point. Using a KDTree algorithm, new points cannot be sampled if they are within a radius equal to $5\%$ of the mosaic's swath width. This ensures that the sampled points are randomly distributed across the mosaic, thus providing a more accurate representation of the benthic cover.

\section{Case Studies}\label{sec:CaseStudies}
\subsection{Locations}
\begin{figure}[!htbp]
\centering
\includegraphics[width=0.99\textwidth]{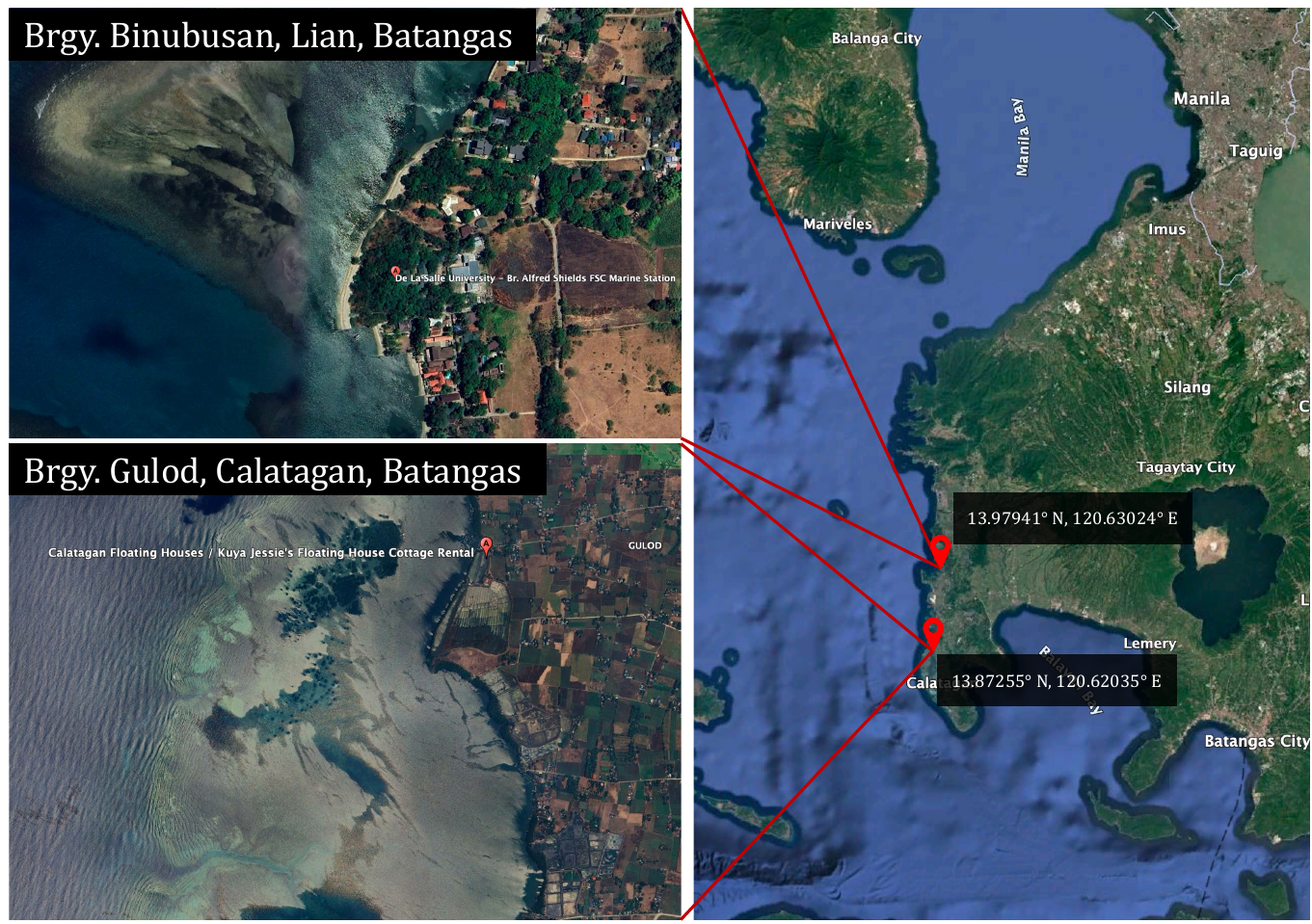}
\caption{The locations of the case study sites. These satellite images were taken from Google Earth Pro.}\label{fig:study_loc}
\end{figure}

To demonstrate the capabilities of CorStitch, we applied it to belt and dive transect videos gathered from two municipalities in Batangas, Philippines. The dive transect video data was collected on May 5, 2025, in Barangay Binubusan, Lian, across $50\,\mathrm{m}$ transects. Meanwhile, belt transect video data were collected in Barangay Gulod, Calatagan, on July 6, 2024, and October 6, 2024 (Figure \ref{fig:study_loc}).

\subsection{Dive transect in Brgy. Binubusan, Lian, Batangas}
\begin{figure}[!htbp]
\centering
\includegraphics[width=0.99\textwidth]{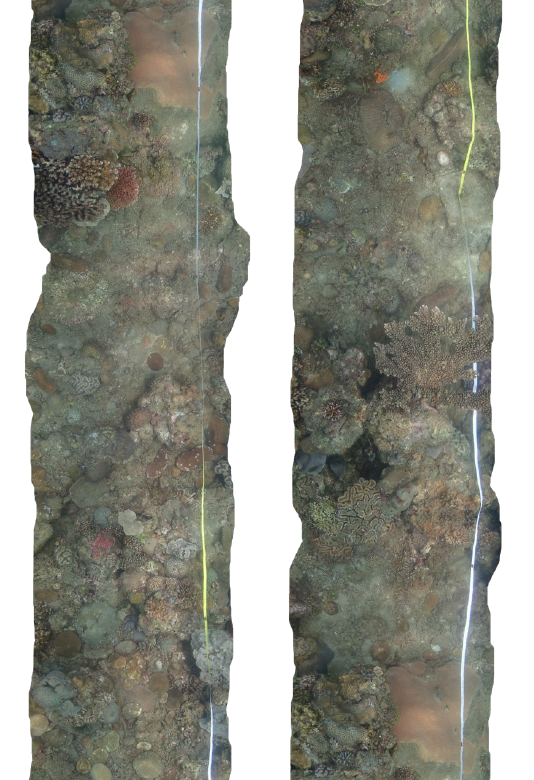}
\caption{The mosaics generated by CorStitch from a dive transect video with a $1$-meter altitude and linear FOV. The two mosaics are sections of one $50$-meter transect, with the transect on the right being a continuation of the transect on the left. The alternating bright and dark bands are caused by the lensing effect of the surface waves. For scale, the black marks on the transect tape have a separation distance of $1$-meter.}\label{fig:1m_linear_dive_transect}
\end{figure}

A diver recorded underwater videos at a depth of $5\,\mathrm{m}$ along a $50$-meter transect by manually holding a down-looking camera. Multiple passes were made along the transect while varying the video collection parameters to demonstrate the flexibility of CorStitch. Shown in Figure \ref{fig:1m_linear_dive_transect} are the mosaics that CorStitch generated from a dive transect captured using the Olympus TG-6 Underwater Camera with a linear FOV while maintaining a $1$-meter altitude from the seabed.  The mosaics are visually seamless, with corals, rocks, sand, and other structures appearing as continuous objects with minimal breaks and misalignments.

As the diver holds the camera, small rotations and changes in elevation can sometimes happen. These non-planar movements can cause stitching artifacts such as small breaks and misalignments. Due to these artifacts, CorStitch is not recommended for coral area measurements. Instead, these mosaics can be used to estimate coral cover and bleaching prevalence. 

\begin{figure}[!htbp]
\centering
\includegraphics[width=0.99\textwidth]{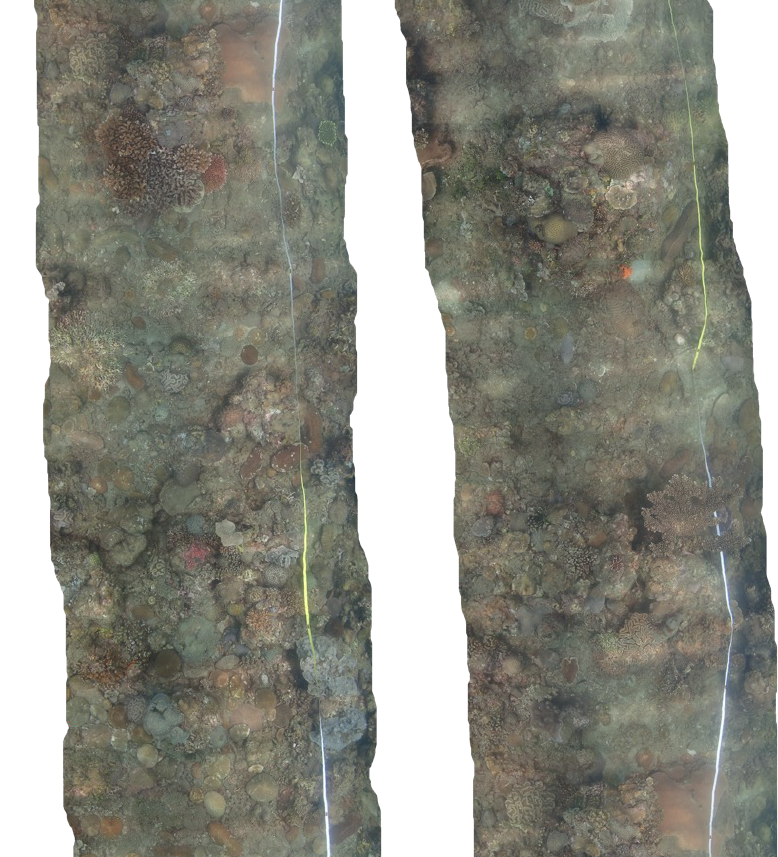}
\caption{The mosaics generated by CorStitch from a dive transect video with $1$-meter altitude and wide-angle FOV. The two mosaics are sections of one $50$-meter transect, with the transect on the right being a continuation of the transect on the left. The alternating bright and dark bands are caused by the lensing effect of the surface waves. For scale, the black marks on the transect tape have a separation distance of $1$-meter.} \label{fig:1m_wide_dive_transect}
\end{figure}

We then tested CorStitch with a wide-angle FOV. The wide-angle has a larger swath width than the linear FOV, but it has a more prominent barrel distortion. Due to the refraction caused by the air-water interface between the camera and the water column, the distortion is lessened compared to in-air images. Comparing the mosaics in Figures \ref{fig:1m_linear_dive_transect} and \ref{fig:1m_wide_dive_transect}, both mosaics are similar in quality. The stitching is visually seamless, and there are minimal breaks and discontinuities in the features.

The choice between linear and wide-angle FOV entirely depends on the user. Although Figures \ref{fig:1m_linear_dive_transect} and \ref{fig:1m_wide_dive_transect} show that the mosaics are similar in image quality and distortions, too much distortion can cause a significant degradation in mosaic quality. To demonstrate this, we collected another video along the same transect using the Sony Alpha 6400 with an ultra-wide-angle lens. The Sony camera's ultra-wide-angle lens produced a larger FOV than the Olympus camera's. The mosaics from the Sony camera are shown in Figure \ref{fig:1m_alpha_wide}. Several corals appear to be stretched or compacted along the transect direction. The features near the border of the swath range are also significantly stretched out due to the barrel distortion, but there is minimal distortion around the center line of the mosaic.

\begin{figure}[!htbp]
\centering
\includegraphics[width=0.99\textwidth]{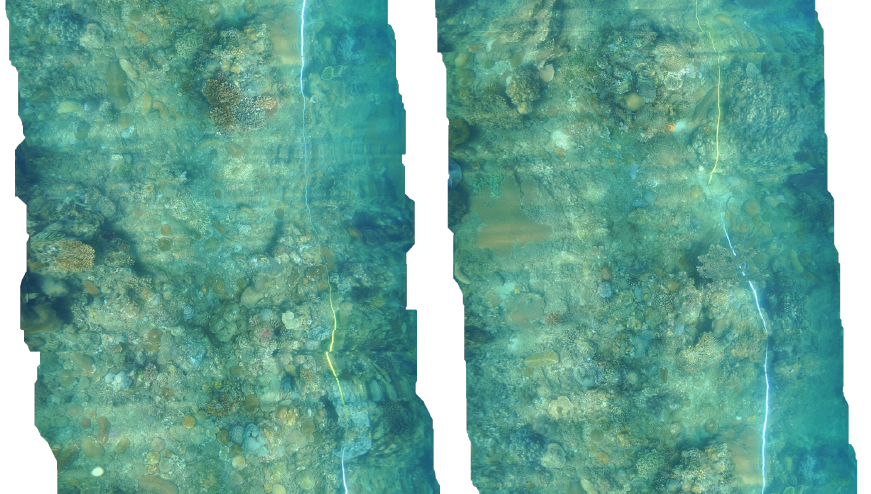}
\caption{The mosaics generated by CorStitch from a dive transect video with $1$-meter altitude and ultrawide-angle FOV. The two mosaics are sections of one $50$-meter transect, with the transect on the right being a continuation of the transect on the left. The alternating bright and dark bands are caused by the lensing effect of the surface waves. For scale, the black marks on the transect tape have a separation distance of $1$-meter. Due to the distortions in the mosaics caused by the ultrawide-angle lens, the scaling is inconsistent throughout the image. }\label{fig:1m_alpha_wide}
\end{figure}

After creating the mosaics, they can be uniformly randomly sampled while maintaining a minimum distance between each sampled point. CorStitch randomly placed red cross marks across the entire mosaic with a minimum separation distance of $5\%$ of the swath width. Each marking is assigned a number for labeling purposes. 
\begin{figure}[!htbp]
\centering
\includegraphics[width=0.99\textwidth]{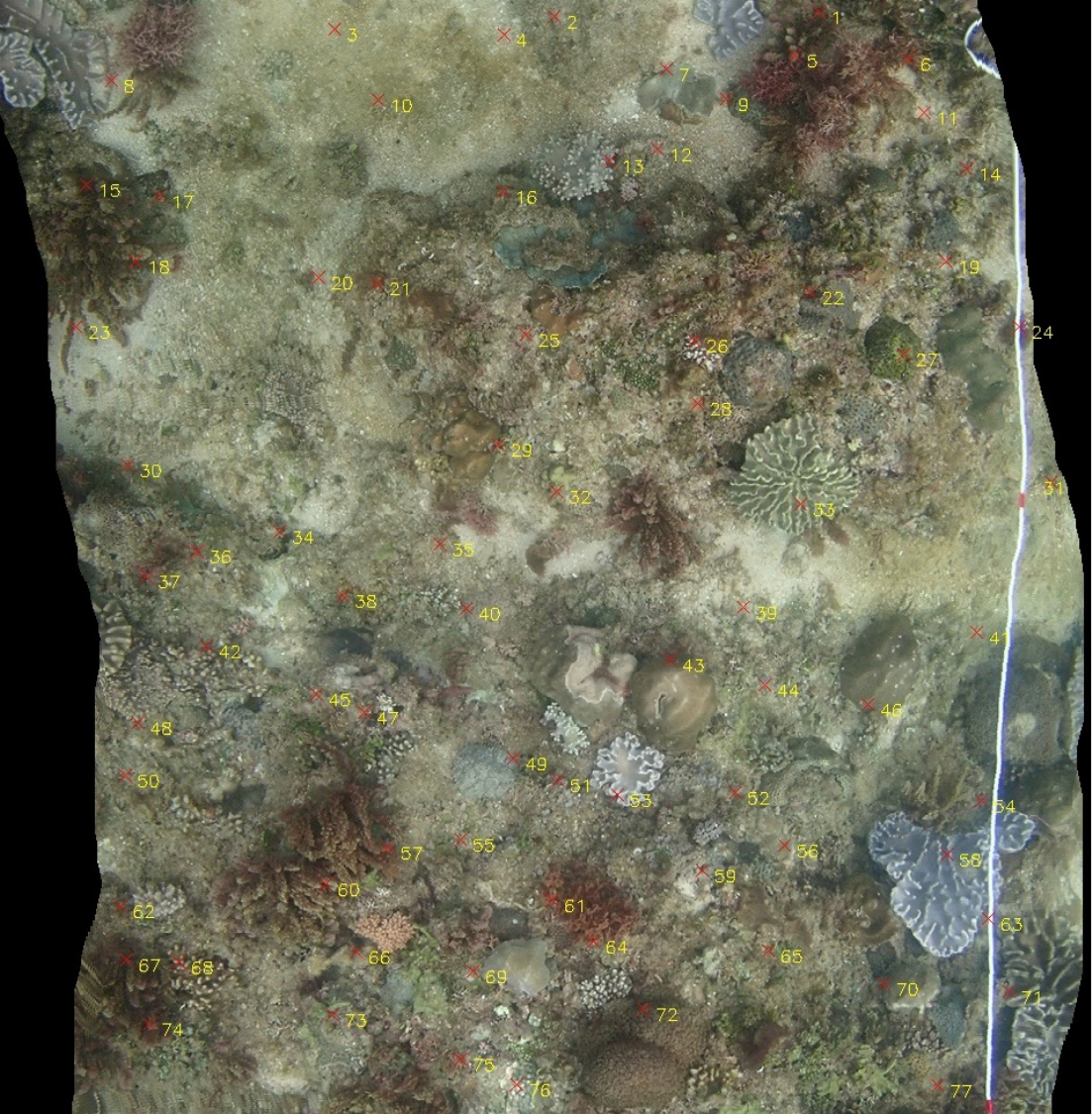}
\caption{A section of the mosaic that was uniformly marked by CorStitch. The markers have a minimum separation distance of $5\%$ of the swath width. The numbering starts at the end of the transect, which is the top of the mosaic. }\label{fig:1m_alpha_wide}
\end{figure}

\newpage
\subsection{Calatagan, Batangas}
Underwater videos from a $4.5\,\mathrm{km}$ belt transect were captured at Brgy. Gulod, Calatagan, Batangas, in July and October 2024. The survey on the two dates followed similar belt transects with some overlaps. We recorded the videos with an ARRAS-type survey using a boat-towable platform with a GoPro Hero 3+ camera and a GPS echosounder attached. The down-facing camera was set to a wide FOV. A Humminbird 899ci HD SI Combo echosounder was used to record the GNSS data and the depth in meters. 

\begin{figure}[!htbp]
\centering
\includegraphics[width=1\textwidth]{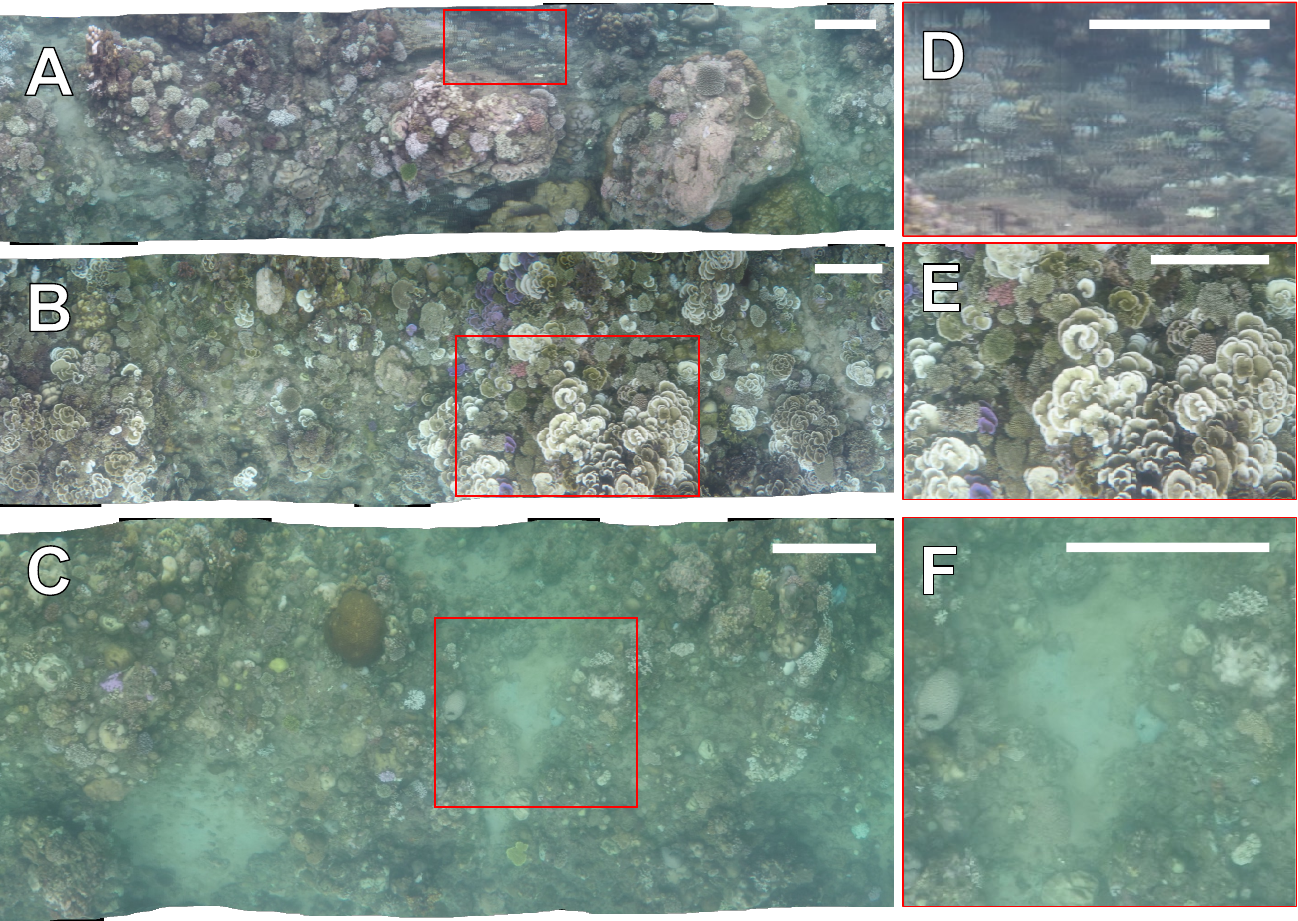}
\caption{Sample reef mosaics generated using CorStitch from Calatagan, Batangas, captured in July 2024. (A) Mosaic showing variations in depth and the presence of towering corals, resulting in (D) stitching misalignment. (B) Mosaic showing a coral reef platform with approximately uniform elevation, resulting in (E) seamless stitching. (C) Mosaic of a deeper coral reef with sand patches, still resulting in (F) seamless stitching. The scale bars correspond to 1 meter, and the transect direction for these mosaics is to the right.}
\label{fig:july2024}
\end{figure}

The recorded videos are approximately 40 minutes long, with the camera towed at an altitude of around $2$ to $4\,\mathrm{m}$ from the seabed.  Figure \ref{fig:july2024} shows the mosaics generated using CorStitch with a resolution of 1080p. We were able to generate 462 mosaics, where each mosaic is 5 seconds' worth of frames.

As observed, CorStitch was able to produce seamless mosaics of successive frames while preserving the distinguishable general morphology of the corals. Figure \ref{fig:july2024}A shows a scene with depth variations containing towering corals. Upon closer inspection, stitching misalignment is evident due to motion blur, as shown in Figure \ref{fig:july2024}D. In contrast, Figure \ref{fig:july2024}B depicts a reef platform with approximately uniform elevation, resulting in seamless stitching, as shown in Figure \ref{fig:july2024}E.

The stitching algorithm identifies corresponding features between successive frames and aligns them based on the strongest feature matches. High-contrast features, such as sharp edges, provide the strongest correlations and are therefore preferentially aligned during stitching. However, motion blur reduces the sharpness of these features, leading to stitching misalignment. This highlights the importance of proper data acquisition. Based on our experiments, we observed that at a depth of around $5\,\mathrm{m}$, a tow speed of approximately $2$ to $4\,\mathrm{m/s}$ minimizes the occurrence of motion blur. At greater depths, higher tow speeds can be used to improve the efficiency of the stitching algorithm while increasing the area covered.

Figure \ref{fig:july2024}C depicts a relatively deeper coral reef with sand patches. Since the camera is white-balancing automatically, the abrupt changes in depth result in green and blue frames, as the camera takes time to adjust to the new scene. While this results in a dominant blue or green mosaic, this will not affect the stitching algorithm, as long as the sharp edges are still present for correlation. As shown in Figure \ref{fig:july2024}F, CorStitch was still able to stitch the large sand patch because there are features surrounding it. 

\begin{figure}[!htbp]
\centering
\includegraphics[width=1\textwidth]{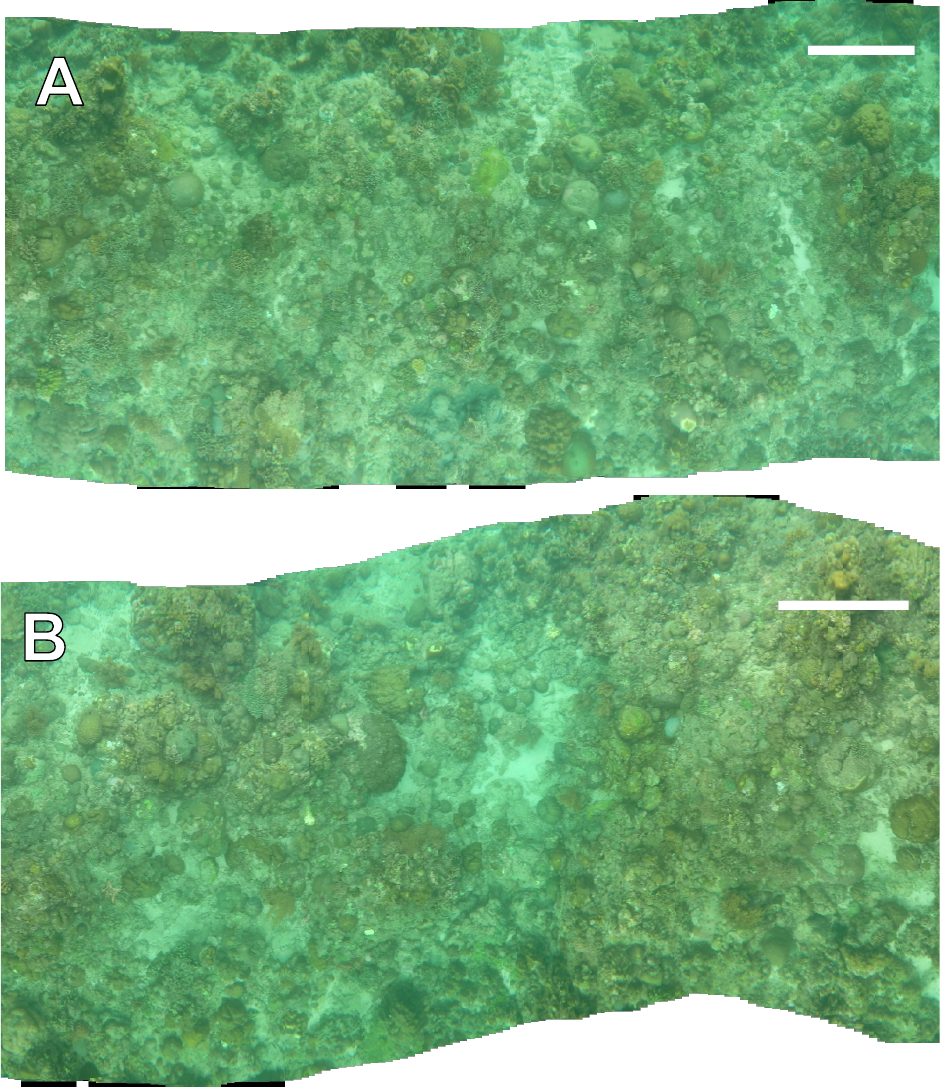}
\caption{Sample reef mosaics generated using CorStitch from Calatagan, Batangas, captured in October 2024. The mosaics are slightly curved due to the camera's across-track movement. The scale bars correspond to 1 meter, and the transect direction for these mosaics is to the right.}
\label{fig:oct2024}
\end{figure}

For the October 2024 dataset, we were able to generate 556 mosaics, with each mosaic being from a 5-second-long clip. Figure \ref{fig:oct2024} shows sample mosaics generated using CorStitch. As observed, the mosaics are slightly curved, which corresponds to the camera's across-track movement. Take note that this is not a pure along-track movement due to environmental interactions such as small rotations and elevations induced by waves and unstable boat movement. Despite these, the mosaic remains visually seamless, which demonstrates the robustness of CorStitch to small rotations and elevation changes in the video.

\begin{figure}[!htbp]
    \centering
    \includegraphics[width=1\linewidth]{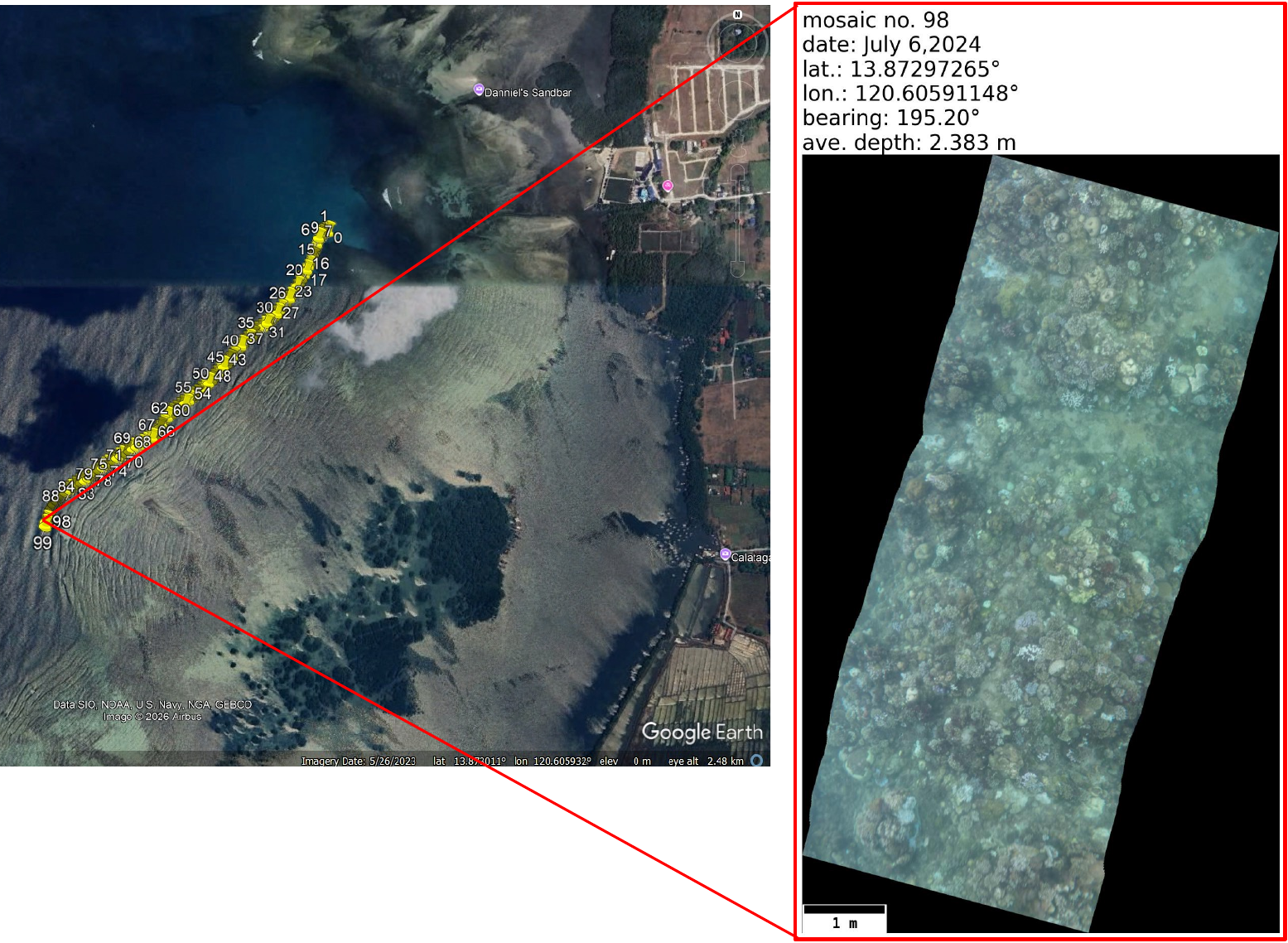}
    \caption{Sample KMZ file of the July dataset showing the generated mosaics from Calatagan, Batangas, viewed in Google Earth. Each pin contains a mosaic oriented according to the calculated bearing, together with the date of data collection, latitude, longitude, bearing, and average depth.}
    \label{fig:rectifiedmosaic}
\end{figure}

We used the GNSS data recorded by the GPS echosounder to georeference the generated mosaics for both the July and October datasets. Figure \ref{fig:rectifiedmosaic} shows a sample KMZ file from the July dataset opened in Google Earth, displaying the first 100 mosaics. Each pin contains a rectified mosaic with information about the date the data were collected, as well as the mosaic's latitude, longitude, bearing, average depth, and a 1-meter scale bar. Each mosaic is oriented based on the calculated bearing, which represents the boat's direction at a given moment.

\begin{figure}[!htbp]
    \centering
    \includegraphics[width=1\linewidth]{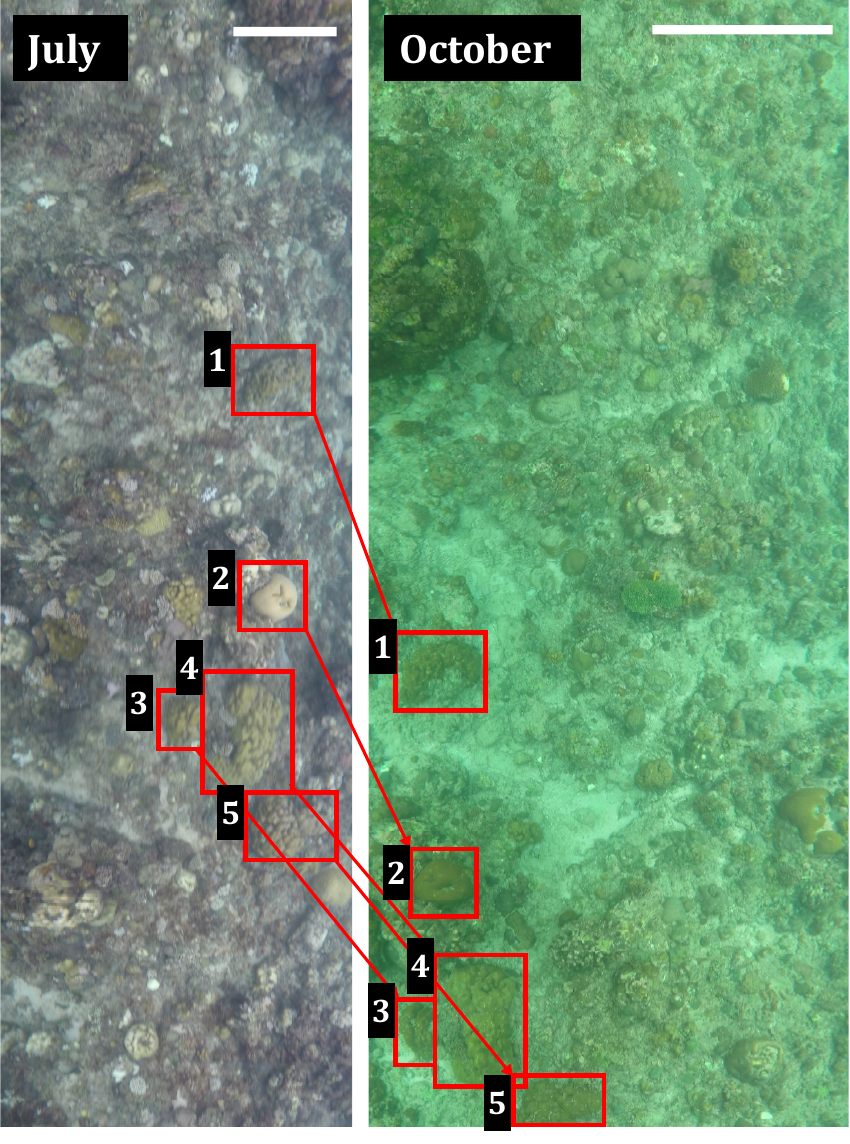}
    \caption{Sample overlapping mosaics from July and October datasets of Calatagan, Batangas. The distinguishable matching corals from the two images are enclosed in red boxes. The average distance between coral pairs is $9\,\mathrm{m}$, measured using Google Earth's ruler feature. The scale bars correspond to 1 meter.}
    \label{fig:matchmosaics}
\end{figure}

Since the survey tracks for both months were nearly identical, we were able to observe overlapping mosaics in several sections. Figure \ref{fig:matchmosaics} shows sample overlapping mosaics containing matching coral colonies. Although the shapes and apparent sizes of the corals may differ due to variations in camera orientation during image acquisition, their overall morphology remains consistent. Using Google Earth's ruler feature, the average distance between the coral pairs in Figure \ref{fig:matchmosaics} is $9\,\mathrm{m}$. Although we have not validated the actual location of these corals, seeing them in the same location on two different dates validates the consistency of CorStitch when synchronizing video timestamps to GNSS coordinates. But even with the average error of $9\,\mathrm{m}$, which could be attributed to the overall accuracy of the GNSS collection system and second-level accuracy of synchronizing the video and GNSS data, the general locations of the corals are still obtained.  This demonstrates the potential of CorStitch as a tool for long-term coral reef monitoring, particularly for assessing changes following bleaching events or natural disasters. Further color correction of the mosaics may improve the visibility of the matching coral colonies.

\newpage
\newpage
\section{CorStitch Features}

\subsection{Image Resolution}

\begin{figure}[!htbp]
\centering
\includegraphics[width=0.98\textwidth]{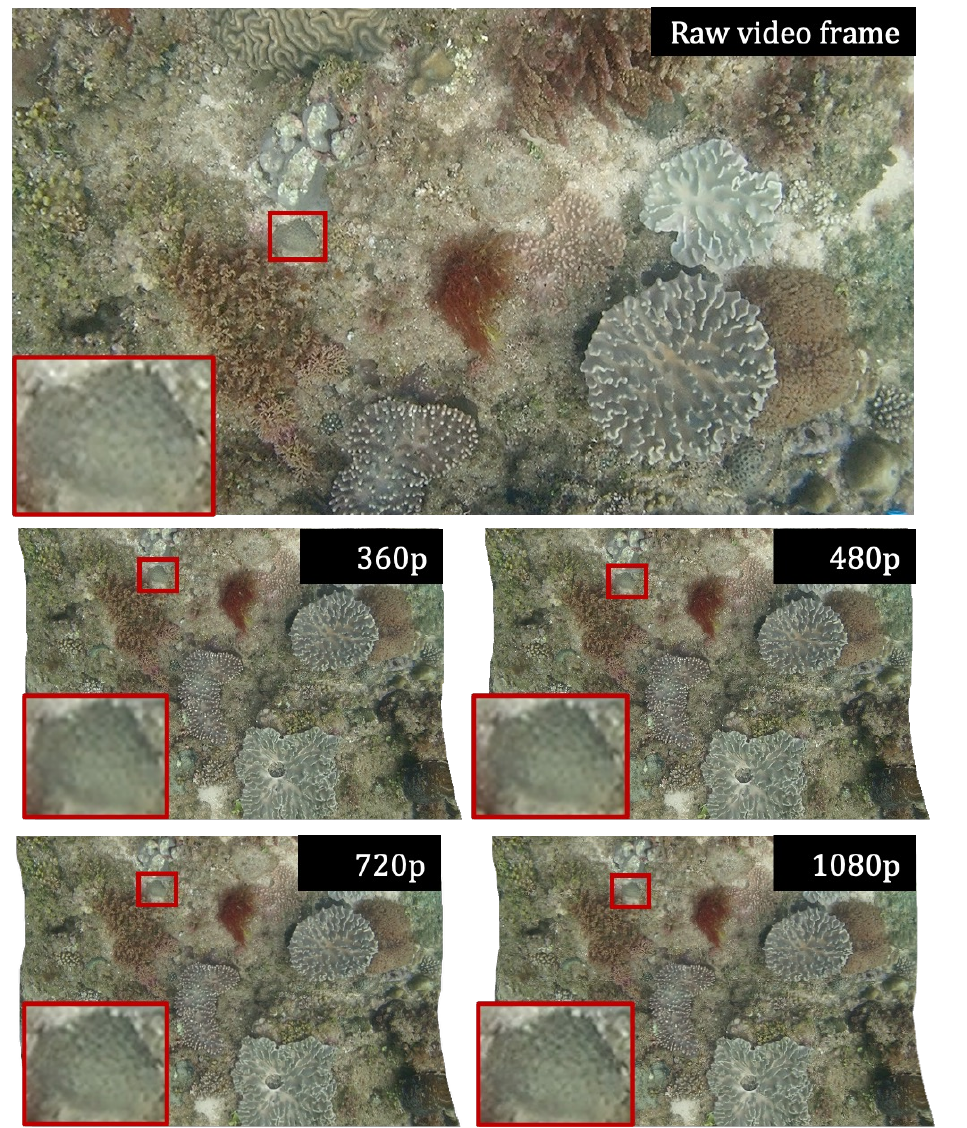}
\caption{A qualitative comparison of a single frame at original video resolution compared with mosaics formed across different resolutions. The inset shows close-ups of a small coral in the middle of the frame.}
\label{fig:quality}
\end{figure}

The mosaics across different resolutions and the raw video frame are shown in Figure \ref{fig:quality}. The raw video frame serves as the highest possible resolution, and it is used as the baseline when qualitatively assessing the mosaics. Across the different resolutions, small details are lost at 360p and 480p. This makes the small boulder coral appear as a rock instead. At 720p and 1080p, the seams of the mosaic are visible, but they do not obstruct the polyps, which allows the proper identification of the coral. During experimentation, we found that 720p and 1080p may be optimal for dive transects because they preserve small details that are crucial for coral identification. For belt transects, 480p serves as a good balance between storage use and mosaic quality.

\subsection{Belt transect: Mosaic time}
The mosaic time dictates the maximum number of frames that CorStitch will use to form the mosaic. However, the pixel length of the mosaic will depend on the correlation between each central strip. This is important for georeferencing belt transects because both depth and tow speed affect the length of the georeferenced mosaics. For a given speed, large depths produce shorter mosaics than shallower depths due to parallax. Additionally, for a given depth, higher tow speeds produce longer mosaics because more information is gathered for a set duration. 

\begin{figure}[!htbp]
\centering
\includegraphics[width=0.98\textwidth]{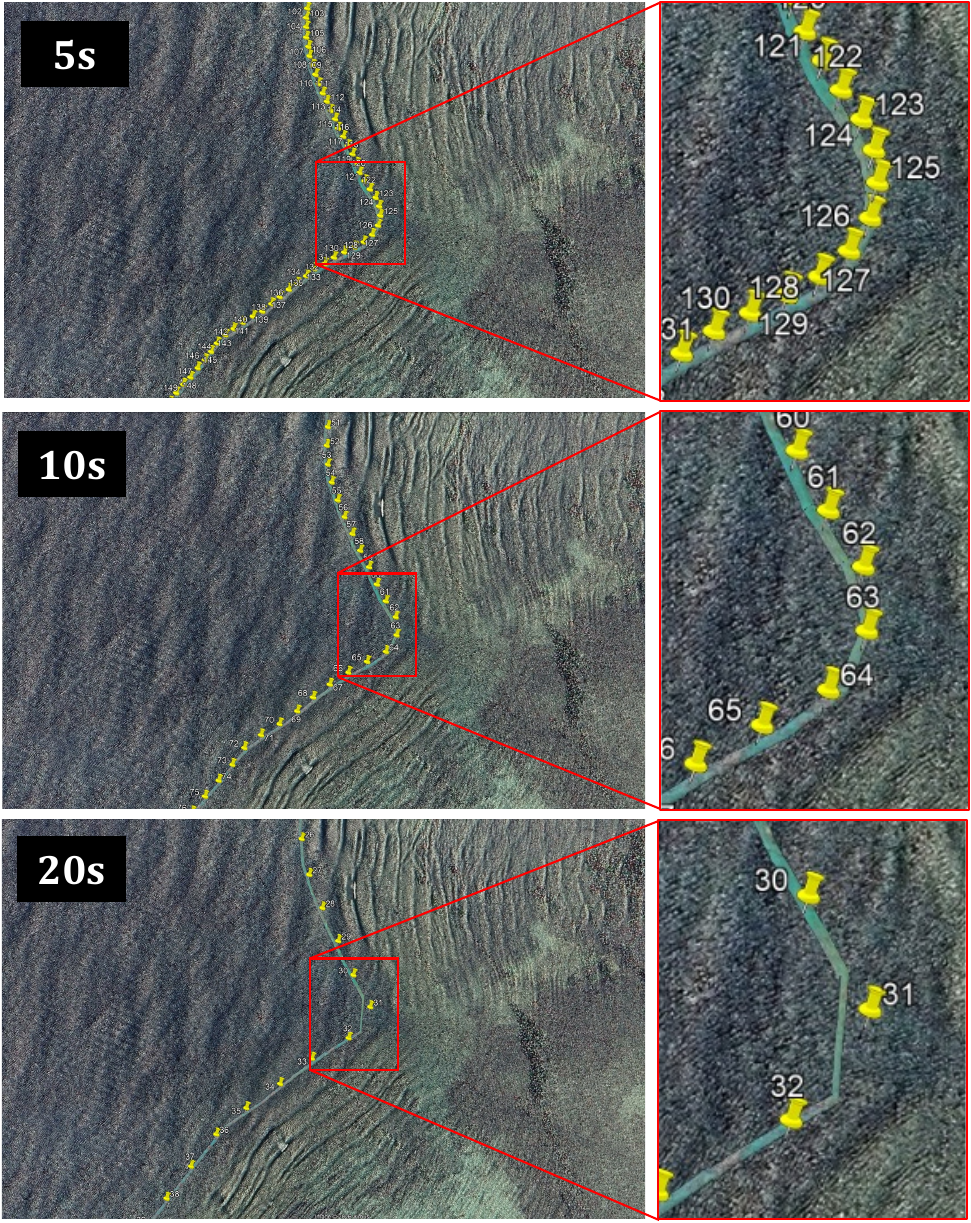}
\caption{Comparison of mosaic duration using $5$-, $10$-, and $20$-second video clips. The mosaics are from the July 2024 dataset of Calatagan, Batangas.}
\label{fig:comparecurve}
\end{figure}

Choosing a mosaic time that is too small may lead to the mosaics becoming too short that may contain insufficient spatial information. If the mosaic time is too long, then the georeferenced mosaic quality will drop because the bearing, latitude, and longitude are only assigned at the corners of the image. To demonstrate this, Figure \ref{fig:comparecurve} shows the effect of mosaic time. For the 5-second mosaic time, the georeferenced mosaics follow a smooth curved path that outlines the surveyed belt transect. For the 10-second mosaics, the number of mosaics is lessened and it can be seen that the curve is still smooth. However, at a 20-second mosaic time, the mosaics become too long and can no longer be georeferenced properly, as evidenced by pin 31 not being aligned with the mosaic. 

\subsection{CorStitch Graphic User Interface}
\begin{figure}[!htbp]
\centering
\includegraphics[width=0.98\textwidth]{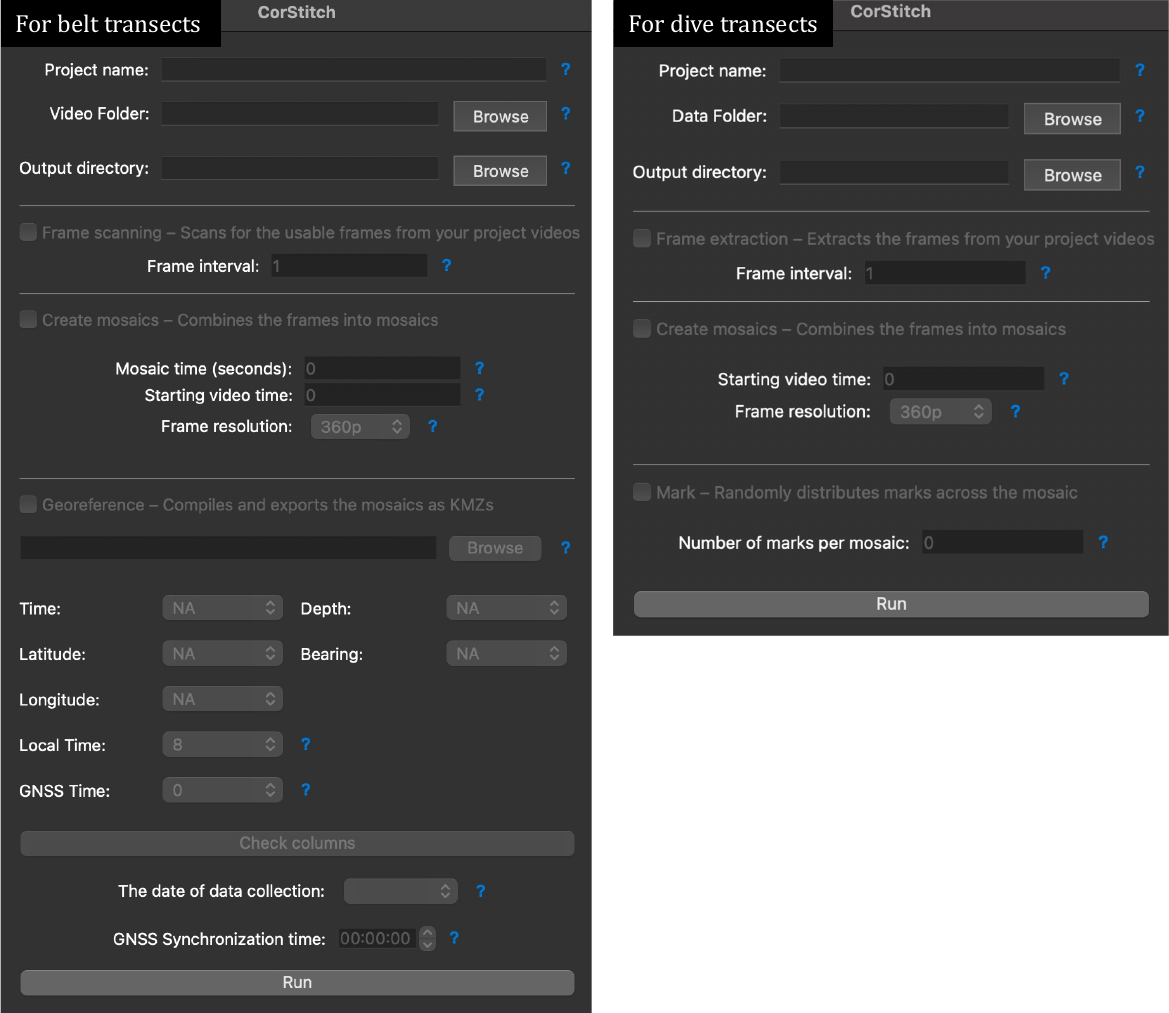}
\caption{The GUI for the belt and dive transect versions of CorStitch. The explanation of each input in the GUI can be found in our manual posted on our GitHub page in section \ref{sec:data}.}
\label{fig:gui}
\end{figure}

To make CorStitch user-friendly, we made a graphic user interface (GUI) for both the belt and dive transect versions of CorStitch shown in \ref{fig:gui}. The GUI collects all the user-needed inputs for processing transect videos. The description of each input and how to run the software are listed in the \href{https://github.com/jclmaypa/CorStitch}{user's manual} of CorStitch.

\section{Conclusion}
Across the two case studies, we have demonstrated that CorStitch can convert both belt and dive transect videos to high-quality mosaics. For dive transects, CorStitch successfully generated mosaics that remained robust to small angular movements and changes in camera elevation. The mosaics are also robust to the distortions introduced by wide-angle lenses but become less reliable when processing videos acquired with ultra-wide-angle fields of view. CorStitch can also uniformly sample random pixels in the mosaic to simulate the sampling method of CPCe, which will allow the user to get the benthic distribution across the mosaic.

For belt transects, we have demonstrated that CorStitch can produce high-quality mosaics despite the presence of motion blur, changes in depth, and small angles caused by the towing of the camera. Since CorStitch uses only the green channel, even predominantly bluish videos were still seamlessly stitched. The georeferenced mosaics show that CorStitch is consistent in synchronizing GNSS data with the mosaic data by showing the same coral colonies around the same area. These results indicate that CorStitch has the potential to support the tracking of temporal changes in benthic distribution.

Although CorStitch is effective for creating underwater video mosaics, it is subject to several limitations arising from its Fourier-based image registration approach. The current implementation assumes that consecutive frames are primarily related by translation and therefore cannot compensate for large rotations, significant changes in camera elevation, and severe parallax. Consequently, motion blur and highly distorted imagery, such as the ultra-wide FOV, may produce stitching artifacts that reduce mosaic quality. In addition, the accuracy of the georeferenced mosaics remains dependent on the precision of the GNSS measurements and the synchronization between the video and navigation data. 

Despite these limitations, CorStitch provides a free, open-source, and user-friendly workflow for generating mosaics from both belt and dive transect videos. By integrating mosaicking, georeferencing, and random point overlay into a single software package, CorStitch reduces the technical and financial barriers associated with underwater video processing. We anticipate that CorStitch will serve as a practical and accessible supplement to existing underwater image-processing workflows for researchers, local communities, government agencies, and non-governmental organizations engaged in long-term coral reef monitoring, assessment, and conservation.

\section{Acknowledgments}
The authors acknowledge the assistance of Jessie and Leo Delos Reyes in the collection of belt transect videos. 

\section{Data}\label{sec:data}
The Python code, the Windows and macOS executables, sample geoereferenced mosaics, and the CorStitch users' manual can be found on our \href{https://github.com/jclmaypa/CorStitch}{Github}.

\bibliography{sn-bibliography}

\end{document}